\pdfoutput=1
\documentclass[11pt]{article}
\usepackage{authblk}
\usepackage{acl2023}
\usepackage{times}
\usepackage{graphicx}
\usepackage{latexsym}
\usepackage{array}
\usepackage{booktabs}
\usepackage{pgfplots}
\usepackage{algorithm}
\usepackage{algpseudocode}
\usepackage{graphicx}
\usepackage[T1]{fontenc}
\usepackage[utf8]{inputenc}
\usepackage{microtype}
\usepackage{xcolor}
\usepackage{soul}\usepackage{times}
\usepackage{graphicx}
\usepackage{latexsym}
\usepackage{array}
\usepackage{booktabs}
\usepackage{pgfplots}
\usepackage{algorithm}
\usepackage{algpseudocode}
\usepackage{graphicx}
\usepackage[T1]{fontenc}
\usepackage[utf8]{inputenc}
\usepackage{microtype}
\usepackage{xcolor}
\usepackage{soul}
\usepackage{times}
\usepackage{latexsym}
\usepackage[T1]{fontenc}
\usepackage[utf8]{inputenc}
\usepackage{microtype}
\usepackage{inconsolata}
\begin{document}
\title{Dense Sparse Retrieval: Using Sparse Language Models for Inference Efficient Dense Retrieval.}
\author[1]{Daniel Campos}
\author[1]{ChengXiang Zhai}
\affil[1]{Department of Computer Science, the University of Illinois Urbana-Champaign}
\maketitle
\begin{abstract}
Vector-based retrieval systems have become a common staple for academic and industrial search applications because they provide a simple and scalable way of extending the search to leverage contextual representations for documents and queries. As these vector-based systems rely on contextual language models,  their usage commonly requires GPUs, which can be expensive and difficult to manage. Given recent advances in introducing sparsity into language models for improved inference efficiency, in this paper, we study how sparse language models can be used for dense retrieval to improve inference efficiency. Using the popular retrieval library Tevatron and the MSMARCO, NQ, and TriviaQA datasets, we find that sparse language models can be used as \textit{direct replacements} with little to no drop in accuracy and up to 4.3x improved inference speeds. 
\end{abstract}
\section{Introduction}
The use of language models for information retrieval has been a well-established field of research for several decades. However, traditional language models rely on dense representations of language, which may result in computational challenges when dealing with large datasets or resource-intensive tasks. To address these challenges, recent research has explored the use of unstructured sparsity in language models, finding it an effective way to improve inference efficiency when paired with sparsity-aware inference frameworks. \\
While pruning models can require specialized knowledge and frameworks, using already pruned models, which can be \textit{sparse-transferred} to new tasks, is attractive as it amortizes compression costs. It allows non-experts to improve inference performance. By using a pruned (sparse) language model, practitioners can realize inference speedups without any additional effort. \\
This paper explores the use of sparse language models for vector-based retrieval. We provide a simple case study demonstrating how the sparse language model, oBERT \cite{Kurti2022TheOB}, can directly replace the commonly used BERT-base  \cite{Devlin2019BERTPO} model. In our experiments on the MSMARCO \cite{Campos2016MSMA}, Natural Questions \cite{Kwiatkowski2019NaturalQA}, and TriviaQA \cite{Joshi2017TriviaQAAL} passage retrieval datasets, we find sparse language models can serve as drop-in replacements for existing models with minor optimization and they deliver up to 4.3x more queries per second (QPS) with little to no loss in accuracy. \\
The contributions of this work are as follows
\begin{itemize}
    \item We demonstrate the use of unstructured sparsity in vector-based retrieval combined with a sparsity aware inference engine to provide up to 4.3x QPS with little to no loss in accuracy.
\end{itemize}
\section{Related Work}
\textbf{Vector-based retrievers}, often called bi-encoders,dual-encoders, or dense retrievers, are retrieval models which leverage an implicit ranking signal from the inner product of query and document representations of contextual language models. They have become incredibly popular as they can be highly accurate \cite{pouran-ben-veyseh-etal-2021-dpr} \cite{Karpukhin2020DensePR} and when paired with an Approximate Nearest Neighbor (ANN) such as FAISS \cite{johnson2019billion}, allow for incredibly efficient retrieval. \\
\textbf{Unstructured Sparsity} is a compression approach in which some portion of a model's weights or groups of weights receive a mask and are effectively removed from the network. Recent work has shown it possible to introduce sparsity into small \cite{Kurti2022TheOB}, \cite{Sanh2020MovementPA}, \cite{Zafrir2021PruneOF} and large language models \cite{Frantar2023SparseGPTML} with little to no loss in accuracy. Using a sparsity-aware inference engine or modern Ampere generation GPUs\footnote{https://developer.nvidia.com/blog/accelerating-inference-with-sparsity-using-ampere-and-tensorrt/}, it is possible to gain 3-5x speedups in inference throughput. While existing research has studied novel tasks where sparsity can perform badly \cite{Liu2023SparsityMC}, transferring to novel domains \cite{Campos2022SparseBERTSM}, to the best of our knowledge, there has been no study in leveraging model sparsity in information retrieval.\\
\section{Leveraging Sparsity For Dense Retrieval}
\subsection{Dense Retrieval Background}
Vector-based retrieval leverages a representation model to create representations for the query and the document in a shared latent space. This can be achieved by using a single model, called a \textit{tied bi-encoder}, or with independent query and document models, called a \textit{untied bi-encoder}. Using these models, representations are made for queries and documents, and a notion of relevance is learned by training to minimize the distance of positive query-document pairs as shown in equation \ref{eq:dis} where $\textbf{x}$ is a query vector and $\textbf{y}$ is a document vector, and $\cdot$ denotes the dot product of the vectors.\\
\begin{equation}
L = 1 - \frac{\textbf{x} \cdot \textbf{y}}{|\textbf{x}||\textbf{y}|}
 \label{eq:dis}
\end{equation}
After models have been trained, an index from the document corpus is created by encoding every document into a vector, and these vectors are loaded into an ANN index. When a query is issued, it encodes into the shared latent space using the query encoder, and the closest documents are retrieved. Since the document index is generated once (or any time an index is refreshed) and offline, it typically leverages batch processing and large accelerators. Conversely, the query encoder must run each time a user query is issued with small batches and commonly on many small query processors, which commonly lack GPU acceleration.\\
\begin{table}[htb!]
      \centering
        {\small 
            \begin{tabular}{l|c}
            \toprule
            Parameter & Possible Values \\
            \midrule
            Training Length& 3,40 Epochs \\
            Initial learning rate & 1e-5, 5e-5, 7e-5, 9e-5\\
            Learning rate schedule &  Linear \\
            \midrule
                Batch size & 8,128, \\
            \midrule
                Negative Passages & 1,8 \\
            \midrule
            \bottomrule
            \end{tabular}
        }
    \caption{Hyperparmaters used to train bi-encoder models for retrieval   \label{tab:hyperparams-sparse-transfer-ir}}
\end{table}
\begin{table*}[htb!]
    \centering
    \caption{Retrieval Accuracy@100 on NQ, MSMARCO, and TriviaQA with respect to inference throughput (queries per second) and relative speedup}
    \begin{tabular}{|l|l|l|l|l|l|}
    \hline
        Model & MSMARCO & NQ & TriviaQA  & QPS & Speedup \\ \hline
        BERT-Base (Pytorch) & 69.80\% & 86.34\% & 85.33\% & 47.278 & 1.00 \\ \hline
        BERT-Base (DeepSparse) & 69.80\% & 86.34\% & 85.33\% & 80.92 & 1.71 \\ \hline
        oBERT 90$\backslash$\% & 70.04\% & 85.84\% & 84.41\% & 202.67 & 4.28  \\ \hline
        oBERT 80$\backslash$\% (block) & 69.63\% & 85.62\% & 84.81\%  & 141.78 & 3.00 \\ \hline
    \end{tabular}
    \label{tab:sum-dense-sparse}
\end{table*}
\subsection{Sparse Transfer Learning}
While it is possible to introduce sparsity in a model using an open-source library like SparseML\footnote{https://github.com/neuralmagic/sparseml}, this can require non-trivial tuning and expertise. Instead, we use already sparsified language models and apply them to new domains. Specifically, we leverage the oBERT 12-layer encoder model \cite{Kurti2022TheOB} and fix the sparsity profile in the new task. The oBERT model has been sparsified during pre-training, and it has a variant with 90\% unstructured sparsity and 80\% block sparsity \cite{Eldar2008BlocksparsityCA}. Using these already sparse models, we explore their usage as drop-in replacements for the uncompressed BERT base.\
\subsection{Experimental Design}
We evaluate the effectiveness of the oBERT models by evaluating how well they can serve as drop-in replacements for BERT-base on various retrieval datasets. We alternate the used model without any major optimization and compare retrieval performance. All of our experiments leverage the open source retrieval library the Tevatron \cite{Gao2022TevatronAE} \footnote{https://github.com/texttron/tevatron} library, which makes use of hugginface's transformers \cite{wolf-etal-2020-transformers}.\\
\textbf{Datasets} We use a wide variety of standard dense retrieval benchmarks, including MSMARCO V1.1 \footnote{https://huggingface.co/datasets/Tevatron/msmarco-passage} \cite{Campos2016MSMA}, NQ \footnote{https://huggingface.co/datasets/Tevatron/wikipedia-nq} \cite{Kwiatkowski2019NaturalQA}, and TriviaQA \footnote{https://huggingface.co/datasets/Tevatron/wikipedia-trivia} \cite{Joshi2017TriviaQAAL} passage retrieval datasets.  \\
For each dataset, we train models to converge with \textit{tied} and \textit{untied} bi-encoders, generate full indexes, and evaluate performance by measuring recall accuracy with retrieval depths of 20,100, and 200. \\
\textbf{Computational Experiments} are all performed on 16 GB V100 GPUS using 1 V100 for MSMARCO and 4 for each other experiment. We use the training hyperparameters found in \ref{tab:hyperparams-sparse-transfer-ir}, finding the sparse language models use higher learning rates consistent with Kurtic et-al. \cite{Kurti2022TheOB} findings.
\subsection{Inference Benchmarks}
We benchmark inference speeds of query encoding to evaluate the impact of using sparse language models. We benchmark using an Intel Xeon Gold 6238R Processor using native Pytorch inference and leverage the sparsity-aware inference library DeepSparse\footnote{https://neuralmagic.com/deepsparse/}. For each variant model, we evaluate the performance on encoding 6500 queries with a batch size of one and a max context length of 32.We repeat each run five times to ensure consistency and report the mean. Detailed results are in the \ref{sec:app-inference}.
\section{Experimental Results}
\begin{figure}[!htb]
\begin{tikzpicture}
\scalebox{0.85}{
\begin{axis}[
    title={Speedup vs. Retrieval Performance},
    xlabel={Queries Per Second},
    ylabel={Retrieval Accuracy @100},
    xmin=75, xmax=250,
    ymin=98.5 , ymax=100.1,
    xtick={100, 150, 200},
    ytick={100,99.75, 99.5, 99.25,99, 98.75},
    legend pos=north east,
    ymajorgrids=true,
    grid style=dashed,
    legend style={nodes={scale=0.4, transform shape}}, 
    legend image post style={mark=*}
]
\addplot[
    color=red,
    mark=square,
    ]
    coordinates {
    (80.92, 100) (141, 99.76) (202.67, 100.0034)
    };
\addplot[
    color=blue,
    mark=square,
    ]
    coordinates {
    (80.92, 100) (141, 99.17) (202.67, 99.43)
    };
\addplot[
    color=green,
    mark=square,
    ]
    coordinates {
    (80.92, 100)  (141, 99.39) (202.67, 98.92)
    };

\legend{MSMARCO, NQ, TriviaQA}
 \end{axis}}
\end{tikzpicture}
    \centering
    \caption{Measuring the impact on recall accuracy at 100 vs. inference throughput on the MSMARCO, NQ, and TriviaQA retrieval datasets \label{fig:sparse}}
\end{figure}
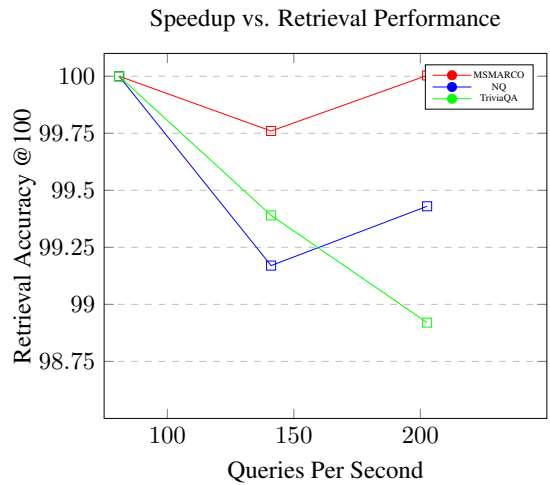

In figure \ref{fig:sparse}, we plot the retrieval performance of retrieval accuracy at 100 relatives to the dense BERT-base baseline finding it is possible to improve inference performance by nearly 4.3x with losing under\% loss in accuracy. Looking at the more detailed results in \ref{tab:sum-dense-sparse}, we find that simply by changing the inference engine from PyTorch to DeepSparse can lead to a 1.7x speedup. Looking at the variation in models, we find that the 90\% sparse model outperforms the 80\% block sparse model regarding inference efficiency and retrieval accuracy. This finding is similar to Kurtic et al., given that block-sparse models see larger inference improvements with quantization, and introducing sparsity in blocks reduces network expressivity. We explored introduced quantization both during training using QAT and post-training, but in both cases, this caused a near-complete collapse of retrieval accuracy.  \\
Looking at the more detailed results found in tables \ref{tab:MSMARCO-Sparse-Tranfer}, \ref{tab:NQ-Sparse-Tranfer}, and \ref{tab:TriviaQA-Sparse-Tranfer}, we can see that the use of sparse language models works well with both tied and untied bi-encoders. Still, performance improves when the bi-encoder is tied, which we attribute to the query and document encoder sharing the less expressive sparse language model. Additionally, the impact of using sparse language models is more pronounced with smaller recall sets. When the recall set is 20, losses range from 1.3\% to 3\%; however, when the recall set is expanded to 100 items, losses range from minor improvements to under 2\%.  
\section{Conclusion and Future Work}
In this work, we have demonstrated how sparse language models can be used for vector-based retrieval. Using the popular MSMARCO, NQ, and TriviaQA passage retrieval dataset, we improve inference efficiency but 4.3x without a ~1\% loss in accuracy. While the focus of this work is narrow in the future, we wish to explore how sparse language models can be combined with quantization and structured pruning to improve performance further. Additionally, we seek to investigate the role of sparsity in smaller models and cross-encoders.  
\bibliography{anthology,custom}
\bibliographystyle{acl_natbib}
\appendix
\section{Dense Retrieval}
\begin{table*}[!htb]
    \centering
    \small
    \scalebox{0.8}{
    \begin{tabular}{|l|l|l|l|l|l|l|l|l|l|}
    \hline
        Model & Tied & Sparsity & Block-Sparsity & Accuracy @20 & Impact & Accuracy @100 & Impact & Accuracy @ 200 & Impact \\ \hline
        BERT-base & Yes & 0.00\% & No & 79.83\% & 0.00\% & 86.34\% & 0.00\% & 88.42\% & 0.00\% \\ \hline
        BERT-base & No & 0.00\% & No & 79.89\% & 0.07\% & 85.82\% & -0.61\% & 88.42\% & 0.00\% \\ \hline
        oBERT & Yes & 90.00\% & No & 78.78\% & -1.32\% & 85.84\% & -0.58\% & 87.70\% & -0.81\% \\ \hline
        oBERT & No & 90.00\% & No & 78.28\% & -1.94\% & 85.71\% & -0.74\% & 87.48\% & -1.07\% \\ \hline
        oBERT & Yes & 80.00\% & Yes & 78.48\% & -1.70\% & 85.62\% & -0.83\% & 87.17\% & -1.41\% \\ \hline
        oBERT & No & 80.00\% & Yes & 78.17\% & -2.08\% & 85.10\% & -1.44\% & 87.17\% & -1.41\% \\ \hline
    \end{tabular}}
    \caption{Performance of sparse-transferred oBERT compared to BERT-base on the NQ passage retrieval dataset}
    \label{tab:NQ-Sparse-Tranfer}
\end{table*}
\begin{table*}[!htb]
    \centering
    \small
    \caption{Performance of sparse-transferred oBERT compared to BERT-base TriviaQA passage retrieval dataset}
    \scalebox{0.8}{
    \begin{tabular}{|l|l|l|l|l|l|l|l|l|l|}
    \hline
        Model & Tied & Sparsity & Block-Sparsity & Accuracy @20 & Impact & Accuracy @100 & Impact & Accuracy @ 200 & Impact \\ \hline
        BERT-base & Yes & 0.00\% & No & 79.62\% & 0.00\% & 85.33\% & 0.00\% & 86.72\% & 0.00\% \\ \hline
        BERT-base & No & 0.00\% & No & 79.43\% & -0.24\% & 85.03\% & -0.35\% & 86.62\% & -0.12\% \\ \hline
        oBERT & Yes & 90.00\% & No & 78.40\% & -1.54\% & 84.41\% & -1.08\% & 86.10\% & -0.72\% \\ \hline
        oBERT & No & 90.00\% & No & 78.29\% & -1.67\% & 84.35\% & -1.14\% & 85.95\% & -0.89\% \\ \hline
        oBERT & Yes & 80.00\% & Yes & 79.31\% & -0.39\% & 84.81\% & -0.62\% & 86.49\% & -0.26\% \\ \hline
        oBERT & No & 80.00\% & Yes & 78.68\% & -1.18\% & 84.60\% & -0.85\% & 86.29\% & -0.50\% \\ \hline
    \end{tabular}}
    \label{tab:TriviaQA-Sparse-Tranfer}
\end{table*}

\begin{table*}[!ht]
    \centering
    \small
    \caption{Performance of sparse-transferred oBERT compared to BERT-base on the MSAMRCO passage retrieval dataset }
    \scalebox{0.7}{
    \begin{tabular}{|l|l|l|l|l|l|l|l|l|l|l|l|}
    \hline
        Model & Tied & Sparsity & Block-Sparsity & MRR@10 & Impact & Accuracy @20 & Impact & Accuracy @100 & Impact & Accuracy @ 200 & Impact \\ \hline
        BERT-base & Yes & 0.00\% & No & 32.20429117 & 0.00\% & 60.21\% & 0.00\% & 69.80\% & 0.00\% & 85.29\% & 0.00\% \\ \hline
        BERT-base & No & 0.00\% & No & 32.09568722 & -0.34\% & 59.84\% & -0.61\% & 68.93\% & 0.00\% & 84.84\% & 0.00\% \\ \hline
        oBERT & Yes & 90.00\% & No & 32.76589007 & 1.74\% & 61.46\% & 2.08\% & 70.04\% & 0.35\% & 85.37\% & 0.10\% \\ \hline
        oBERT & No & 90.00\% & No & 30.93550166 & -3.94\% & 58.83\% & -2.30\% & 67.94\% & -1.44\% & 83.68\% & -1.36\% \\ \hline
        oBERT & Yes & 80.00\% & Yes & 32.51235958 & 0.96\% & 60.69\% & 0.79\% & 69.63\% & -0.25\% & 84.50\% & -0.92\% \\ \hline
        oBERT & No & 80.00\% & Yes & 31.24041479 & -2.99\% & 59.18\% & -1.71\% & 68.02\% & -1.32\% & 83.19\% & -1.94\% \\ \hline
    \end{tabular}}
    \label{tab:MSMARCO-Sparse-Tranfer}
\end{table*}
\section{Inference Benchmarks}
\label{sec:app-inference}
Evaluation of inference for the BERT-base model using pytorch can be found in \ref{tab:bert-cpu}, BERT-base model using DeepSparse can be found in \ref{tab:bert-deepsparse}  oBERT 90\% in \ref{tab:obert90}, and oBERT 80\% (block) in \ref{tab:obert80}. 

\begin{table*}[!ht]
    \centering
    \caption{Inference Benchmark for BERT-Base using Pytorch}
    \small
    \scalebox{0.8}{
    \begin{tabular}{|l|l|l|l|l|l|l|l|l|}
    \hline
        items/sec & Full Time & Mean Time & 95th & 50th & 5th & 99th & 75th \\ \hline
        Run 1 & 44.890 & 80.414 & 2.17E-02 & 2.92E-02 & 2.09E-02 & 1.97E-02 & 3.07E-02 & 2.21E-02 \\ \hline
        Run 2 & 48.370 & 74.628 & 2.01E-02 & 2.11E-02 & 2.00E-02 & 1.96E-02 & 2.22E-02 & 2.03E-02 \\ \hline
        Run 3 & 47.290 & 76.334 & 2.06E-02 & 2.19E-02 & 2.04E-02 & 1.96E-02 & 2.28E-02 & 2.11E-02 \\ \hline
        Run 4 & 48.260 & 74.810 & 2.01E-02 & 2.13E-02 & 2.00E-02 & 1.95E-02 & 2.22E-02 & 2.04E-02 \\ \hline
        Run 5 & 47.580 & 75.872 & 2.04E-02 & 2.14E-02 & 2.03E-02 & 1.98E-02 & 2.28E-02 & 2.07E-02 \\ \hline
        average & 47.278 & 76.412 & 2.06E-02 & 2.30E-02 & 2.03E-02 & 1.96E-02 & 2.41E-02 & 2.09E-02 \\ \hline
        stdev & 1.410 & 2.348 & 6.46E-04 & 3.49E-03 & 3.65E-04 & 1.04E-04 & 3.68E-03 & 7.20E-04 \\ \hline
        CI & 1.236 & 2.058 & 5.66E-04 & 3.06E-03 & 3.20E-04 & 9.14E-05 & 3.23E-03 & 6.31E-04 \\ \hline
        Lower & 46.042 & 74.353 & 2.00E-02 & 1.99E-02 & 2.00E-02 & 1.96E-02 & 2.09E-02 & 2.03E-02 \\ \hline
        High & 48.514 & 78.470 & 2.12E-02 & 2.60E-02 & 2.06E-02 & 1.97E-02 & 2.74E-02 & 2.15E-02 \\ \hline
    \end{tabular}}
    \label{tab:bert-cpu}
\end{table*}

\begin{table*}[!ht]
    \centering
    \caption{Inference Benchmark for BERT-Base using DeepSparse}
    \small
    \scalebox{0.8}{
    \begin{tabular}{|l|l|l|l|l|l|l|l|l|}
    \hline
        Run & items/sec & Full Time & Mean Time & 95th & 50th & 5th & 99th & 75th \\ \hline
        1 & 86.03 & 41.96 & 1.09E-02 & 1.16E-02 & 1.11E-02 & 1.02E-02 & 1.19E-02 & 1.13E-02 \\ \hline
        2 & 85.64 & 42.15 & 1.10E-02 & 1.22E-02 & 1.10E-02 & 1.02E-02 & 1.30E-02 & 1.12E-02 \\ \hline
        3 & 83.66 & 43.15 & 1.13E-02 & 1.37E-02 & 1.12E-02 & 1.01E-02 & 1.38E-02 & 1.14E-02 \\ \hline
        4 & 82.23 & 43.90 & 1.14E-02 & 1.34E-02 & 1.14E-02 & 1.02E-02 & 1.60E-02 & 1.17E-02 \\ \hline
        5 & 67.03 & 53.85 & 1.42E-02 & 1.60E-02 & 1.45E-02 & 1.14E-02 & 1.61E-02 & 1.57E-02 \\ \hline
        average & 80.92 & 45.00 & 1.18E-02 & 1.34E-02 & 1.18E-02 & 1.04E-02 & 1.42E-02 & 1.23E-02 \\ \hline
        stdev & 7.91 & 5.01 & 1.35E-03 & 1.67E-03 & 1.52E-03 & 5.65E-04 & 1.88E-03 & 1.95E-03 \\ \hline
        CI & 6.94 & 4.39 & 1.19E-03 & 1.46E-03 & 1.33E-03 & 4.95E-04 & 1.65E-03 & 1.71E-03 \\ \hline
        Lower & 73.98 & 40.62 & 1.06E-02 & 1.19E-02 & 1.05E-02 & 9.94E-03 & 1.25E-02 & 1.06E-02 \\ \hline
        High & 87.86 & 49.39 & 1.29E-02 & 1.48E-02 & 1.32E-02 & 1.09E-02 & 1.58E-02 & 1.40E-02 \\ \hline
    \end{tabular}}
    \label{tab:bert-deepsparse}
\end{table*}

\begin{table*}[!ht]
    \centering
    \caption{Inference Benchmark for oBERT 90\% using DeepSparse}
    \small
    \scalebox{0.8}{
    \begin{tabular}{|l|l|l|l|l|l|l|l|l|}
    \hline
        Run & items/sec & Full Time & Mean Time & 95th & 50th & 5th & 99th & 75th \\ \hline
        1 & 190.31 & 18.97 & 4.67E-03 & 4.52E-03 & 4.24E-03 & 4.20E-03 & 1.88E-02 & 4.29E-03 \\ \hline
        2 & 205.59 & 17.56 & 4.26E-03 & 4.33E-03 & 4.22E-03 & 4.19E-03 & 4.83E-03 & 4.24E-03 \\ \hline
        3 & 204.52 & 17.65 & 4.28E-03 & 4.33E-03 & 4.24E-03 & 4.21E-03 & 5.01E-03 & 4.25E-03 \\ \hline
        4 & 205.11 & 17.60 & 4.27E-03 & 4.34E-03 & 4.23E-03 & 4.20E-03 & 4.82E-03 & 4.24E-03 \\ \hline
        5 & 207.80 & 17.37 & 4.21E-03 & 4.25E-03 & 4.16E-03 & 4.13E-03 & 4.52E-03 & 4.18E-03 \\ \hline
        average & 202.67 & 17.83 & 4.34E-03 & 4.35E-03 & 4.22E-03 & 4.19E-03 & 7.60E-03 & 4.24E-03 \\ \hline
        stdev & 7.02 & 0.65 & 1.89E-04 & 9.91E-05 & 3.28E-05 & 3.16E-05 & 6.29E-03 & 3.99E-05 \\ \hline
        CI & 6.15 & 0.57 & 1.66E-04 & 8.69E-05 & 2.87E-05 & 2.77E-05 & 5.51E-03 & 3.49E-05 \\ \hline
        Lower & 196.51 & 17.27 & 4.17E-03 & 4.27E-03 & 4.19E-03 & 4.16E-03 & 2.10E-03 & 4.20E-03 \\ \hline
        High & 208.82 & 18.40 & 4.50E-03 & 4.44E-03 & 4.25E-03 & 4.21E-03 & 1.31E-02 & 4.27E-03 \\ \hline
    \end{tabular}}
    \label{tab:obert90}
\end{table*}

\begin{table*}[!ht]
    \centering
    \caption{Inference Benchmark for oBERT 80\% (Block) using DeepSparse}
    \small
    \scalebox{0.8}{
    \begin{tabular}{|l|l|l|l|l|l|l|l|l|}
    \hline
        Run & items/sec & Full Time & Mean Time & 95th & 50th & 5th & 99th & 75th \\ \hline
        1 & 134.14 & 26.91 & 6.84E-03 & 6.71E-03 & 0.00622558593750:00 & 6.16E-03 & 1.81E-02 & 6.32E-03 \\ \hline
        2 & 141.06 & 25.59 & 6.44E-03 & 6.51E-03 & 6.19E-03 & 6.14E-03 & 9.44E-03 & 6.25E-03 \\ \hline
        3 & 146.08 & 24.71 & 6.19E-03 & 6.28E-03 & 6.16E-03 & 6.12E-03 & 6.42E-03 & 6.19E-03 \\ \hline
        4 & 143.82 & 25.10 & 6.30E-03 & 6.35E-03 & 6.22E-03 & 6.17E-03 & 6.55E-03 & 6.25E-03 \\ \hline
        5 & 143.80 & 25.10 & 6.31E-03 & 6.37E-03 & 6.26E-03 & 6.21E-03 & 6.56E-03 & 6.28E-03 \\ \hline
        average & 141.78 & 25.48 & 6.42E-03 & 6.44E-03 & 6.21E-03 & 6.16E-03 & 9.42E-03 & 6.26E-03 \\ \hline
        stdev & 4.63 & 0.86 & 2.52E-04 & 1.71E-04 & 4.00E-05 & 3.55E-05 & 5.03E-03 & 4.74E-05 \\ \hline
        CI & 4.06 & 0.75 & 2.21E-04 & 1.50E-04 & 3.50E-05 & 3.11E-05 & 4.41E-03 & 4.15E-05 \\ \hline
        Lower & 137.72 & 24.73 & 6.20E-03 & 6.29E-03 & 6.17E-03 & 6.13E-03 & 5.01E-03 & 6.21E-03 \\ \hline
        High & 145.84 & 26.24 & 6.64E-03 & 6.59E-03 & 6.24E-03 & 6.19E-03 & 1.38E-02 & 6.30E-03 \\ \hline
    \end{tabular}}
    \label{tab:obert80}
\end{table*}
\end{document}